\documentclass[pdflatex,sn-nature]{sn-jnl}

\usepackage{graphicx}%
\usepackage{multirow}%
\usepackage{amsmath,amssymb,amsfonts}%
\usepackage{amsthm}%
\usepackage{mathrsfs}%
\usepackage[title]{appendix}%
\usepackage{xcolor}%
\usepackage{textcomp}%
\usepackage{manyfoot}%
\usepackage{booktabs}%
\usepackage{algorithm}%
\usepackage{algorithmicx}%
\usepackage{algpseudocode}%
\usepackage{listings}%
\usepackage{soul}%
\usepackage[export]{adjustbox}
\usepackage{indentfirst}
\usepackage{anyfontsize}
\begin{document}

\title{All-dielectric Metaphotonics for Advanced THz Control of Spins} 

%All-dielectric metasurface for THz control of spins in a ferrimagnetic iron garnet}


\author*[1]{\fnm{Lucas} \sur{van Gerven}}\email{lucas.vangerven@ru.nl}
\equalcont{These authors contributed equally to this work.}

\author[2,3]{\fnm{Daria O.} \sur{Ignatyeva}}\email{daria.ignatyeva@gmail.com}
\equalcont{These authors contributed equally to this work.}

\author[2,3]{\fnm{Daniil V.} \sur{Konkov}}\email{konkov.dv19@physics.msu.ru}

\author[1]{\fnm{V.} \sur{Bilyk}}\email{vladislav.bilyk@ru.nl}

\author[1]{\fnm{T.} \sur{Metzger}}\email{thomas.metzger@ru.nl}

\author[2,3]{\fnm{Denis M.} \sur{Krichevsky}}\email{dm.krichevsky@physics.msu.ru }

\author[3]{\fnm{Svetlana A.} \sur{Evstigneeva}}\email{konkov.dv19@physics.msu.ru}

\author[3]{\fnm{Petr M.} \sur{Vetoshko}}\email{pvetoshko@mail.ru}

\author[2,3]{\fnm{Vladimir I.} \sur{Belotelov}}\email{belotelov@physics.msu.ru}

\author[1]{\fnm{Aleksei V.} \sur{Kimel}}\email{
aleksei.kimel@ru.nl}

\affil[1]{Institute for Molecules and Materials, 6525 AJ Nijmegen, The Netherlands}

\affil[2]{\orgdiv{Faculty of Physics}, \orgname{Lomonosov Moscow State University}, \orgaddress{\street{Leninskie Gory, 1, bld. 2}, \city{Moscow}, \postcode{119992}, \country{Russia}}}

\affil[3]{\orgname{Russian Quantum Center}, \orgaddress{\street{Skolkovo Innovation Center, Bolshoi Blv., 30, bl.1}, \city{Moscow}, \postcode{121205},\country{Russia}}}


\abstract{While nearly single cycle THz pulse is conventionally accepted as the stimulus for the fastest and the most energy efficient control of spins in magnets, all-dielectric metasurfaces have been recently demonstrated to be the least dissipative mean to enhance and control the coupling of light to spins. All-dielectric metasurfaces for the THz control of spins hold great potential in the field of spintronics and related technologies, pushing the boundaries of speed and energy efficiency in spin-based information processing. Here we demonstrate such a metasurface for an advanced THz control of spins in a ferrimagnetic film of iron garnet. Structuring a nonmagnetic substrate one can force a THz electromagnetic field, otherwise described by plane waves, to acquire an out-of-plane magnetic field and thus enable arbitrary direction of the torque acting on spins in all three dimensions. Hence, metaphotonics opens up a plethora of opportunities for advanced control of spins at THz rates in many hot fields of contemporary science, including spintronics, magnonics and quantum computing.}




\maketitle

During the last decade, metaphotonics has had a significant impact on spin control in magnetic materials. In particular, plasmonic antennas have been used in the cutting-edge technology of Heat Assisted Magnetic Recording~\cite{pan2009heat}, facilitating magnetic writing with the help of light at the nanoscale well below the diffraction limit. These plasmonic antennas can enhance electromagnetic near-fields and opto-magnonic interactions in a broad spectral range from visible light to THz radiation~\cite{zhang2023generation,fan2025spatiotemporal,lee2025enhanced}. In particular, the antennas designed for the THz spectral range boosted the effect of a nearly single cycle THz pulses on the spins in orthoferrites and allowed to reveal the temporal and spectral fingerprints of all-coherent spin switching in $\mathrm{TmFeO_3}$~\cite{schlauderer2019temporal}.

Recently, an alternative approach for an efficient coupling of light to spins was proposed using the framework of all-dielectric metaphotonics. In particular, optomagnonic cavities formed by magnetic spheres~\cite{haigh2015magneto,almpanis2020spherical,rameshti2022cavity} or photonic crystals surrounding a magnetic film~\cite{pantazopoulos2019high,klos2014photonic} was shown to enhance the coupling of visible and near-infrared light to spins. All-dielectric magnetic metasurfaces obtained by patterning highly transparent material with a large refraction index also demonstrated similar functionalities Ref.~\cite{ignatyeva2022all,qin2022nanophotonic}. Compared to plasmonic antennas, such structures exhibit a much higher transparency and resonances with a superior quality-factor. These features enhance the efficiency of light-matter coupling and reduce the energy losses. All-dielectric metasurfaces were shown to exhibit resonances of guided or localized modes~\cite{bsawmaii2022magnetic,gamet2017enhancement,barsukova2017magneto}, which result in unusual magneto-optical behavior~\cite{ignatyeva2020all, xia2022circular} and non-trivial spin dynamics~\cite{chernov2020all,ignatyeva2024optical}.

%Up to now the all-dielectric metasurfaces has been applied for the spin control by light. However, the unique metasurface features may allow one to achieve great advances in the THz spectral range as well, making THz electric and magnetic fields much stronger compared to the state-of-the-art plasmonics~\cite{pancaldi2024terahertz,koya2023advances}. That will be of particular value as 

Ultrafast THz spin control and its practical applications, including Joule-loss-free optomagnonic data processing devices and quantum technologies \cite{walowski2016perspective}, would greatly benefit from such an all-dielectric metasurface for the THz spectral range. Aiming to understand the mechanisms and efficiency of the coupling of THz pulses to spins in real metasurfaces, we perform computational study, design and fabricate an all-dielectric metasurface allowing to couple nearly single cycle THz pulses to a spin resonance in a thin film of iron garnet. We experimentally show that the metasurface not only enhances the coupling, but also substantially changes the electromagnetic fields of the otherwise freely propagating plane wave. These changes generate new, previously absent components of the electromagnetic field and thus excite spin resonances in the ferrimagnetic iron garnet, which a plane wave would not be able to excite. 

%\section*{All-dielectric magnetic metasurface}\label{sec2}

To fabricate a metasurface for THz control of spins we took a 3.6~$\mu$m thick epitaxial film of ferrimagnetic $\mathrm{(BiGd)_3Fe_5O_{12}}$ 
bismuth-substituted rare-earth iron garnet (BIG) grown on a 510~$\mu$m thick paramagnetic gadolinium-scandium gallium garnet (GSGG) substrate (Fig. 1a) (Methods). The film has "easy plane" type of magnetic anisotropy. Applying a rather weak in-plane magnetic field above 0.5~mT turns the film into a single-domain state (Supplementary, Fig. S1). Magnetic properties of the BIG-film are normally modelled in terms of 
three magnetic sublatticies: two antiferromagnetically coupled sublattices of Fe$^{3+}$ ions in octahedral and tetrahedral positions and one sublattice of Gd$^{3+}$ in dodecahedral positions. The magnetization of Gd$^{3+}$ couples antiferromagnetically to the net magnetization of Fe$^{3+}$. THz magnetic fields can launch in such ferrimagnets a mode of spin resonance corresponding to coupled motion of spins of rare-earth and Fe$^{3+}$ ions ~\cite{blank2021thz}.

Performing all-optical experiments similar to Ref.~\cite{parchenko2016non} (Methods) we estimated the frequency of this mode for our garnet as  $f_\mathrm{exc}=0.31 \pm 0.04$THz and its spectral width as 0.08 THz (Fig. 1b). 

\begin{figure}[ht]
\centering
{\bf a}~~~~~~~~~~~~~~~~~~~~~~~~~~~~~~~~~~~~~~~~~~~~~~~~~~~~~~~~~~~~~~~{\bf b}~~~~~~~~~~~~~~~~~~~~~~~~~~~~~~~~~~~~~~~~~~~~~~~~~~~~~~~~~~~~~~~~~~~\\ \includegraphics[width=0.45\linewidth,valign=t]{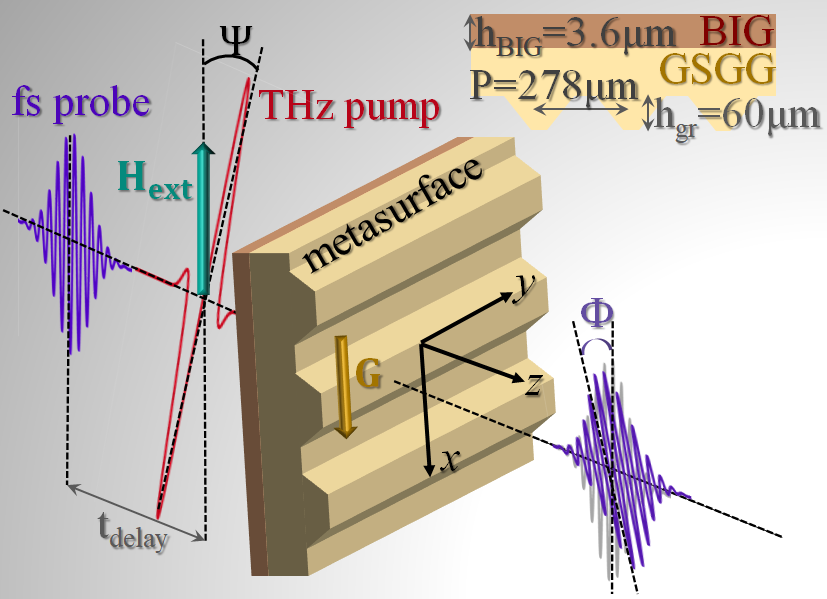}
\includegraphics[width=0.54\linewidth,valign=t]{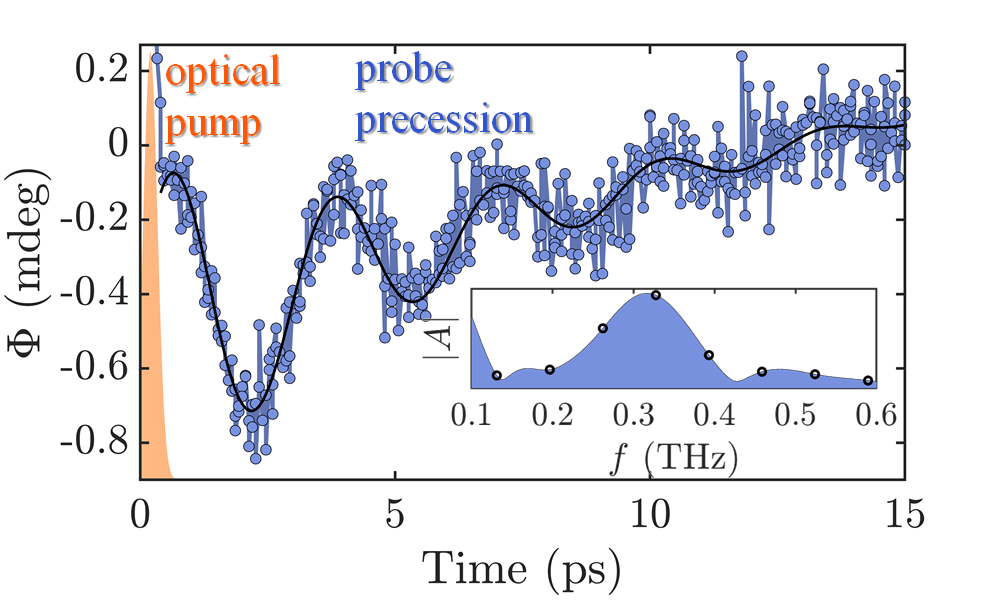}\\
{\bf c}~~~~~~~~~~~~~~~~~~~~~~~~~~~~~~~~~~~~~~~~~~~~~~~~~~~~~~~~~~~~~~~~~~~~~{\bf d}~~~~~~~~~~~~~~~~~~~~~~~~~~~~~~~~~~~~~~~~~~~~~~~~~~~~~~~\\
\includegraphics[width=0.55\linewidth,valign=b]{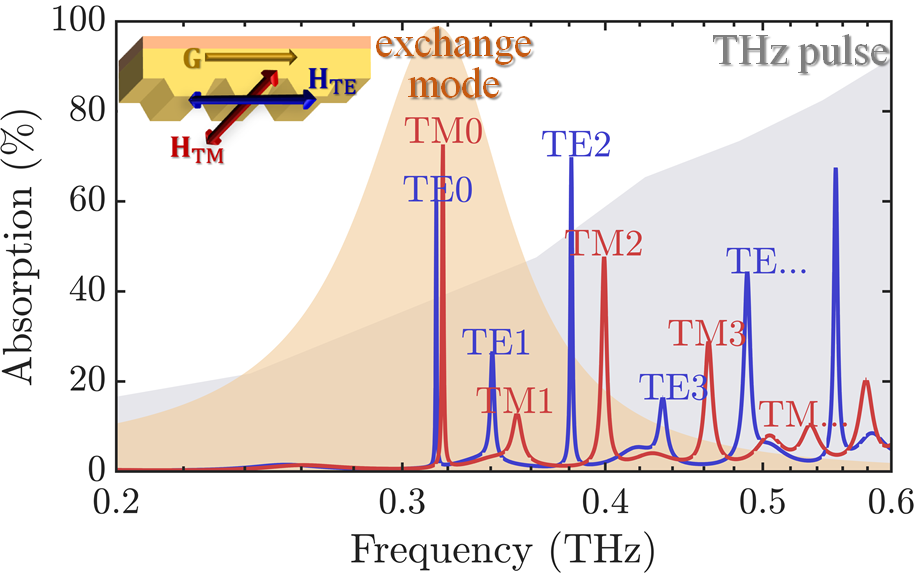}
\begin{minipage}[b]{0.44\linewidth}
\includegraphics[width=0.99\linewidth,valign=b]{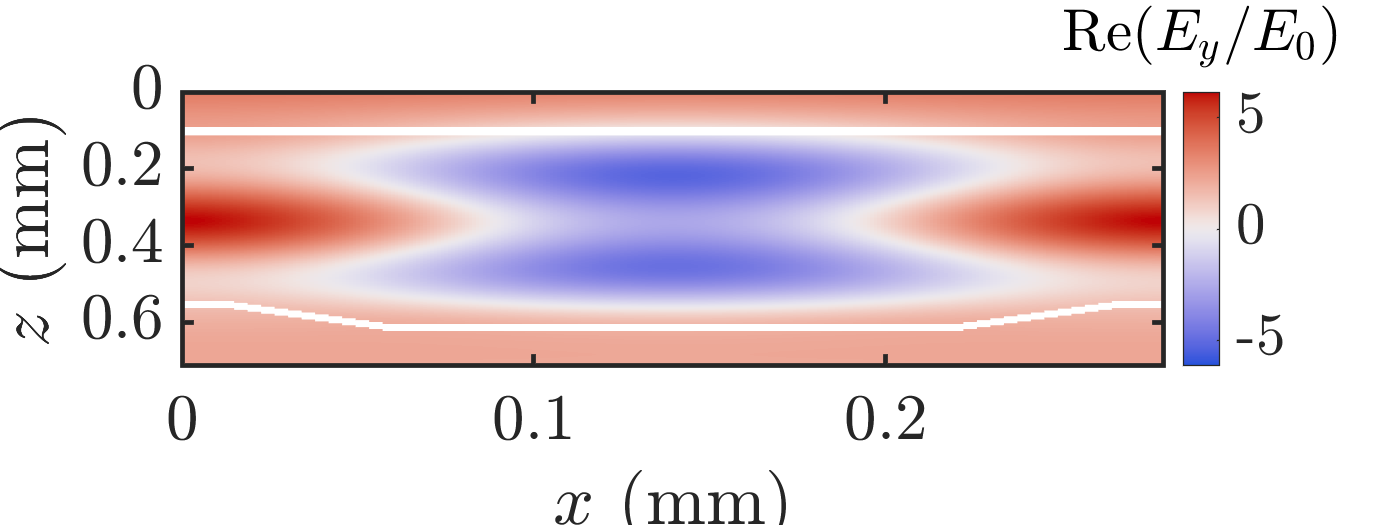}\\
{\bf e}\\
\includegraphics[width=0.99\linewidth,valign=b]{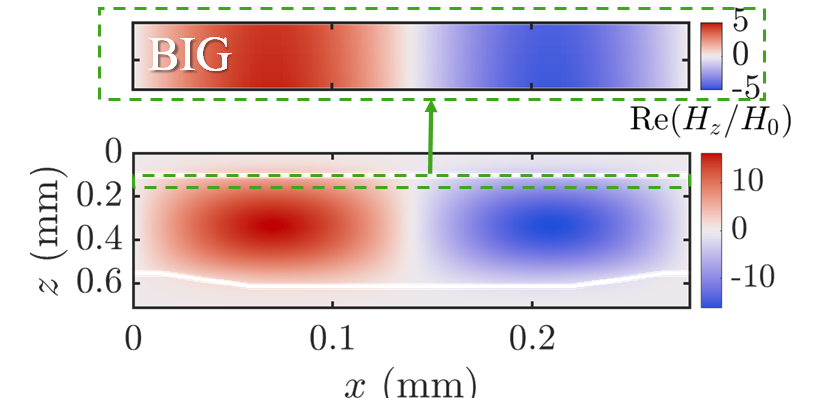}
\end{minipage}

\caption{{\bf THz metasurface for spin control in a (BiGd)$_3$Fe$_5$O$_{12}$ (BIG) film.} {\bf(a)} Schematics of the experiment. {\bf(b)} Transient magneto-optical Faraday effect (dots) induced by a femtosecond laser pulse reveals oscillations corresponding to the exchange resonance mode between the spins of Gd$^{3+}$ and Fe$^{3+}$
ions. The line is a fit with a damped sinusoidal function at the frequency $f=$0.31 THz. The inset shows the Fourier transform of the transient. {\bf(c)} Metasurface absorption spectra demonstrating sharp guided-mode resonances (see the legends). {\bf(d)} Distribution of $E_y$ component (real part) in the metasurface and BIG film (indicated by white lines) at $f=0.315$~THz corresponding to TE0-mode. One period of the metasurface is shown. The distribution reveals the TE0-mode is clearly present. {\bf(e)} Distribution of the TE0-mode induced $Re(H_z)$ in the metasurface and BIG film. The inset zooms in the area of the BIG film.} \label{Fig: General}
\end{figure}

The all-dielectric magnetic metasurface was designed in a form of 1D subwavelength grating of tranches etched by a photolithography (Methods) inside the GSGG substrate (Fig. 1a and Supplementary) by the depth of $60~\mu$m thus leaving $450~\mu$m thick unpatterned GSGG adjacent to the BIG film. The average trench width is $50~\mu$m while the period of the metasurface was chosen to be $278~\mu$m. The metasurface grating is described by the grating vector $\mathbf{G}$ which is  perpendicular to its trenches and $|\mathbf{G}|=2\pi/P$, where $P$ is the period of the metasurface (Fig.~\ref{Fig: General}a). 

There are two types of the guided modes in the sample: the transverse electric (TE) and transverse magnetic (TM) modes propagating perpendicular to the metasurface trenches. They are excited by the incident THz pulse due to the 1D grating which provides phase matching of the fields of the incident electromagnetic wave and the guided mode. The guided modes  are identified as a comb of sharp peaks in the absorption spectrum of the sample for two orthogonal polarizations of the THz pulse: the TE(TM) modes are observed for the magnetic field of the THz pulse $\mathbf{H}_\mathrm{THz}$ perpendicular(parallel) to the grating tranches, i.e. if $\mathbf{H}_\mathrm{THz} || \mathbf{G}$ ($\mathbf{H}_\mathrm{THz} \perp \mathbf{G}$) (see inset and blue and red spectra, respectively, in Fig.~\ref{Fig: General}c).      
There is a family of guided modes TE$N$(TM$N$) of different orders $N$ defined by a number of nodes along the thickness of the waveguide layer. The spectral position of the guided mode resonances $f_{N}^{TE(TM)}$ is found from their dispersion $f_{N}^{TE(TM)}(k)$, and phase matching conditions $\mathbf{k}_{N}^{TE(TM)} = m \mathbf{G}$, where $\mathbf{k}_{N}^{TE(TM)}$ is the mode wavevector and $m$ is a diffraction order exciting the mode. The metasurface period was calculated to match the frequency of the TE0 mode $f_{TE0}$ with the exchange spin resonance in the BIG-film $f_\mathrm{exc}= 0.31$~THz. It is seen from Fig.~\ref{Fig: General}c, that the relatively broad spectral line of the spin resonance also covers TE1-mode as well as TM0- and TM1-modes and even some modes of higher orders, however, the efficiency of their excitation is notably smaller since they already appear at the fringes of the spectral line.   

Figure~\ref{Fig: General}d shows that the distribution of the component $E_y$ of the TE0 mode is nearly homogeneous in the $z$ direction (with $N=0$ nodes of $E_y$) and change sign in the $x$ direction (with $m=1$ phase changes by $2\pi$ per structure period), as expected for a waveguide mode of the zeroth order $N=0$ excited by the first order of diffraction $m=1$. 

The most striking effect of the metasurface is the generation of a normal component of the oscillating magnetic field $H_z$ inside the metasurface, which is absent in the incident electromagnetic field (Fig.~\ref{Fig: General}e). The field is generated as a result of excitation of the TE0-mode in the metasurface, and it is seen that the $H_z$ field can even be boosted to a level 5 times higher than the amplitude of the incident magnetic field (see the upper inset in Fig.~\ref{Fig: General}e).

%\subsection*{Excitation of spin resonance by THz pulses in conventional and unusual field configurations}\label{sec3}

In order to demonstrate the effects of the metasurface on THz-induced spin dynamics and the role of the normal component of the field $H_z$, we employ a pump-probe technique as described elsewhere~\cite{blank2021thz}.  We excited the BIG film on the patterned substrate with a nearly single cycle THz pulse of electromagnetic field (pump) with amplitude up to 1 MV/cm. The spectrum of the pulse was centered at 0.7 THz (Supplementary, Fig. S3). The induced spin dynamics was probed using a pulse with duration of 100 fs at the central wavelength of 800 nm. Upon propagation through the sample the linearly polarized probe pulse experienced polarization rotation due to the magneto-optical Faraday effect. Hence, the Faraday effect measured as a function of pump-probe delay revealed transient spin dynamics induced by the THz pulses in the sample (Methods).

\begin{figure}[ht]

\includegraphics[width=1\textwidth,valign=t]{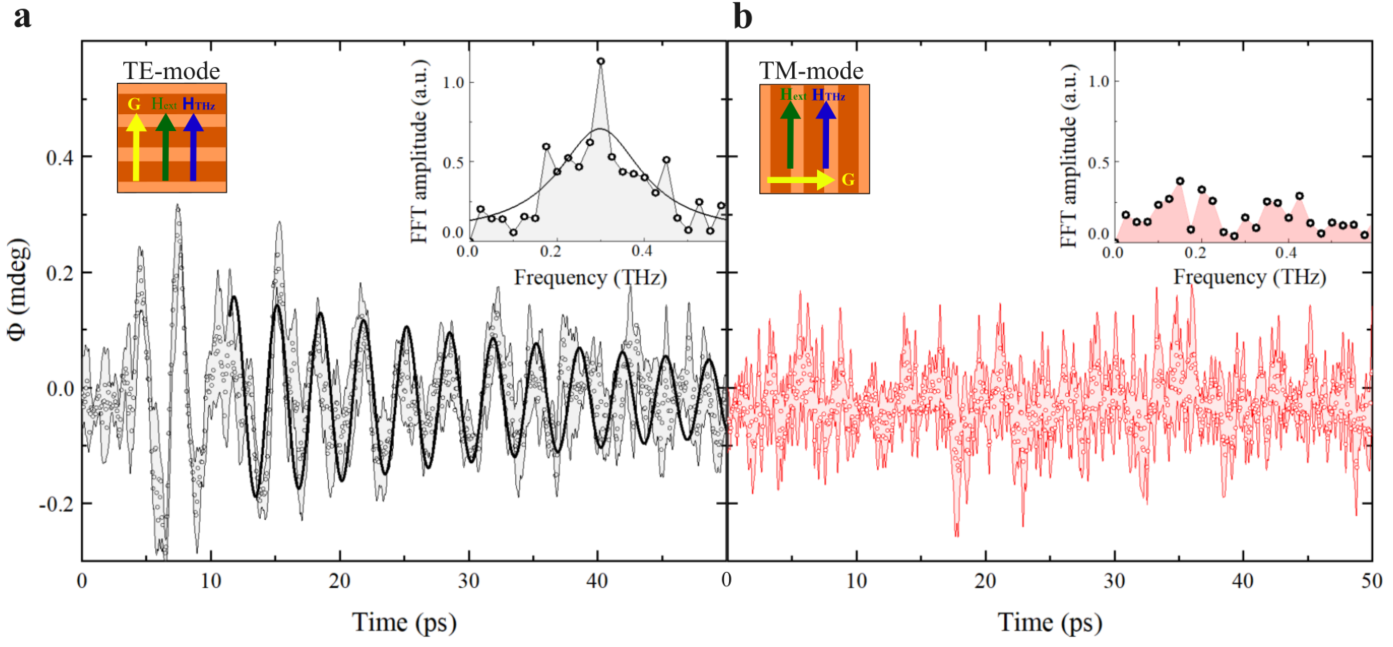}\\
\caption{ \textbf{THz-induced dynamics of the magneto-optical Faraday effect obtained with the help of metasurface for THz spin control in a (BiGd)$_3$Fe$_5$O$_{12}$ (BIG) film.} Transient Faraday rotation induced upon excitation of the TM-mode (panel (\textbf{a})) and the TE-mode (panel (\textbf{b})) in the metasurface. The schematics in the left corners of the panels show mutual orientations of the THz magnetic field $\bf{H}_{THz}$, the external magnetic field $\bf{H}_{ext}$ and the grating vector $\bf{G}$ in the experiments. Solid black curve in (\textbf{a}) is approximation by decaying sine function at $f_\mathrm{exc}=0.31$THz. Inset in the right corners are Fourier transforms of the corresponding transients.}
\label{Fig: Thz signals}
\end{figure}

Figure~\ref{Fig: Thz signals}a shows THz induced dynamics, when the THz electromagnetic field excites the TE0-mode. In this case the THz magnetic field is perpendicular to the trenches  ($\mathbf{H}_\mathrm{THz}||\mathbf{G}$) . The experiments are performed in an external magnetic field, which saturates magnetization in the sample plane ($H_\mathrm{ext}= 100$~mT) aligning it along the THz magnetic field ($\mathbf{H}_\mathrm{ext}||\mathbf{H}_\mathrm{THz}$). This fact excludes any torque of the THz magnetic field on the magnetization ($[\mathbf{H}_{THz} \times \bf {M}] =0$). Despite this fact, the experimental results shown in the figure do reveal pronounced dynamics. It suggests that interacting with the metasurface THz electromagnetic field acquires components perpendicular to the magnetization $\bf {M}$ and magnetic field along the $z$-axis, in particular. Rotating the metasurface by 90-degrees around the $z$-axis and repeating the same experiment with $\mathbf{H}_\mathrm{THz}\perp\mathbf{G}$ and $\mathbf{H}_\mathrm{ext}||\mathbf{H}_\mathrm{THz}$, i.e. under conditions when the THz electromagnetic field does not excite the TE0-mode, we do not observe any THz induced spin dynamics (see Fig. \ref{Fig: Thz signals}b). The observations strongly suggest that the THz-induced  dynamics shown in  Fig. \ref{Fig: Thz signals}a are due to excitation of the TE0-mode and generation of the normal component of the THz magnetic field $H_z$, in particular. 

In order to obtain deeper insights into the observed THz induced spin dynamics, we suggest to focus on a time window, where one can neglect the propagation effects of pump pulses reflected in the studied structure (see Refs.~\cite{grishunin2019transient,subkhangulov2016terahertz}).  In particular, these effects are substantially weaker after the pump pulse had enough time to travel through the studied structure twice. We assume that the substrate (Gd,Sc)$_3$Ga$_5$O$_{12}$ has the very same refraction coefficient in the THz frequency range as the parent compound Gd$_3$Ga$_5$O$_{12}$  ($n_\mathrm{GGG}=3.5$ at $f=0.313$~THz). Having the total thickness of the sample of the order of 500 $\mu$m suggests that roughly after the first 10 ps the propagation effects can be neglected. Hence we perform Fast Fourier Transform (FFT) of the dynamics shown in Fig.~\ref{Fig: Thz signals}a in a window starting from 10 ps after the sample was excited by a pump pulse. The FFT spectrum is shown in Fig.~\ref{Fig: Thz signals}a and reveals a peak at $f= 0.31$ THz that agrees with the frequencies of the TE0-mode in the metasurface and the frequency of the spin resonance in the BIG film. Hence, we conclude that the metasurface generates out-of-plane component of the electromagnetic field $H_z$ which was absent in the incident plane wave, and thereby efficiently excites spin oscillations in the ferrimagnetic BIG film.

 Interestingly, if one excites in the metasurface a TM0-mode ($\mathbf{H}_\mathrm{THz}\perp\mathbf{H}_\mathrm{ext}||\mathbf{G}$), the spin dynamics can also be launched due to the $y$-component of the THz magnetic field $H_y$. It is indeed the case in our experiment, where the spin dynamics induced by the TM0-mode is found to have a similar amplitude as the one for the case of spin dynamics excited by the TE0-mode (Supplementary, Fig. S4). At the same time, we did not observe any THz induced signals from the unpatterned area of the studied sample (Supplementary, Fig. S5). Regarding the level of noise in our measurements, we can conclude that the metasurface enhances the $y$-component of the THz magnetic field $H_y$ at least by a factor of 2. Consequently, the $z$-component of the THz magnetic field induced by the metasurface must be of the same order as the enhanced $H_y$ field. This finding agrees with the initial simulations suggesting that the metasurface will boost the THz magnetic field by a factor of 5. 

\begin{figure}[ht]
\centering
{\bf a}~~~~~~~~~~~~~~~~~~~~~~~~~~~~~~~~~~~~~~~~~~~~~~~~~~~~~~~~~~~~~~~~~~~~~~~~~~~~~~~~~~~~~~~~~~~~~~~~~~~~~~~~~~~~~~~\\
\includegraphics[width=0.98\textwidth,valign=t]{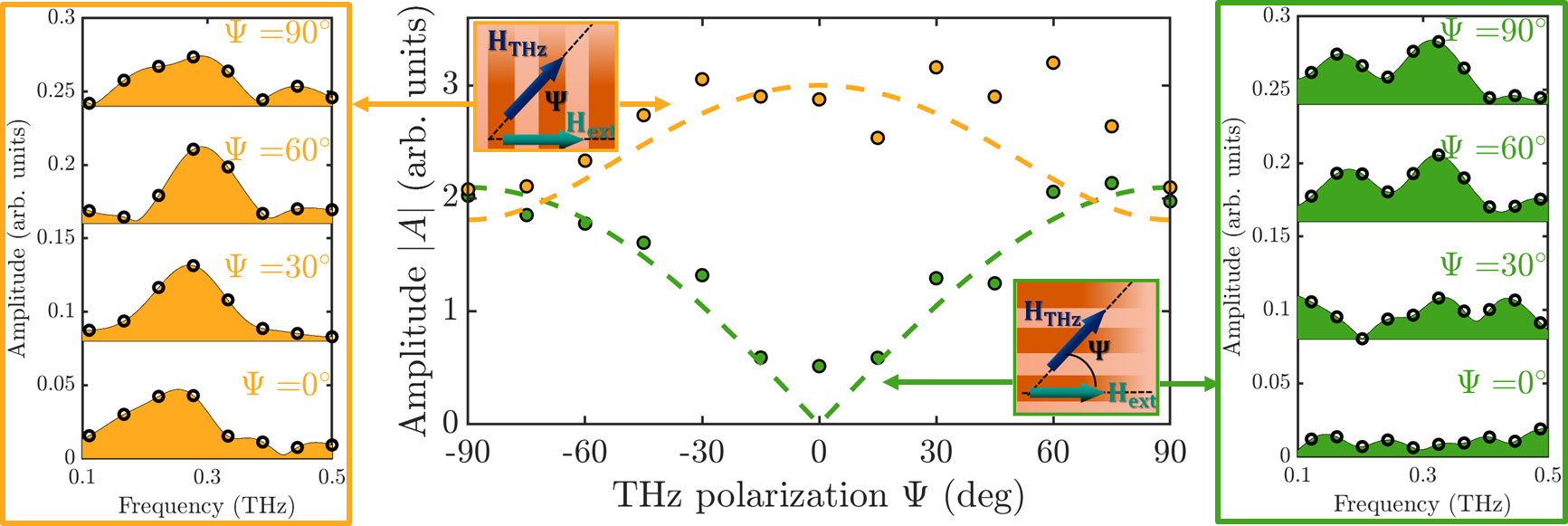}\\
~~~~~~~~~~~~~~~~~~~~~~{\bf b}~~~~~~~~~~~~~~~~~~~~~~~~~~~~~~~~~~~~~~~~~~~~~~~~~~~~~~~~~~~~~~~~~~~~~~~~~~~~~~~~~~~~~~~~~\\
\includegraphics[width=0.55\textwidth,valign=t]{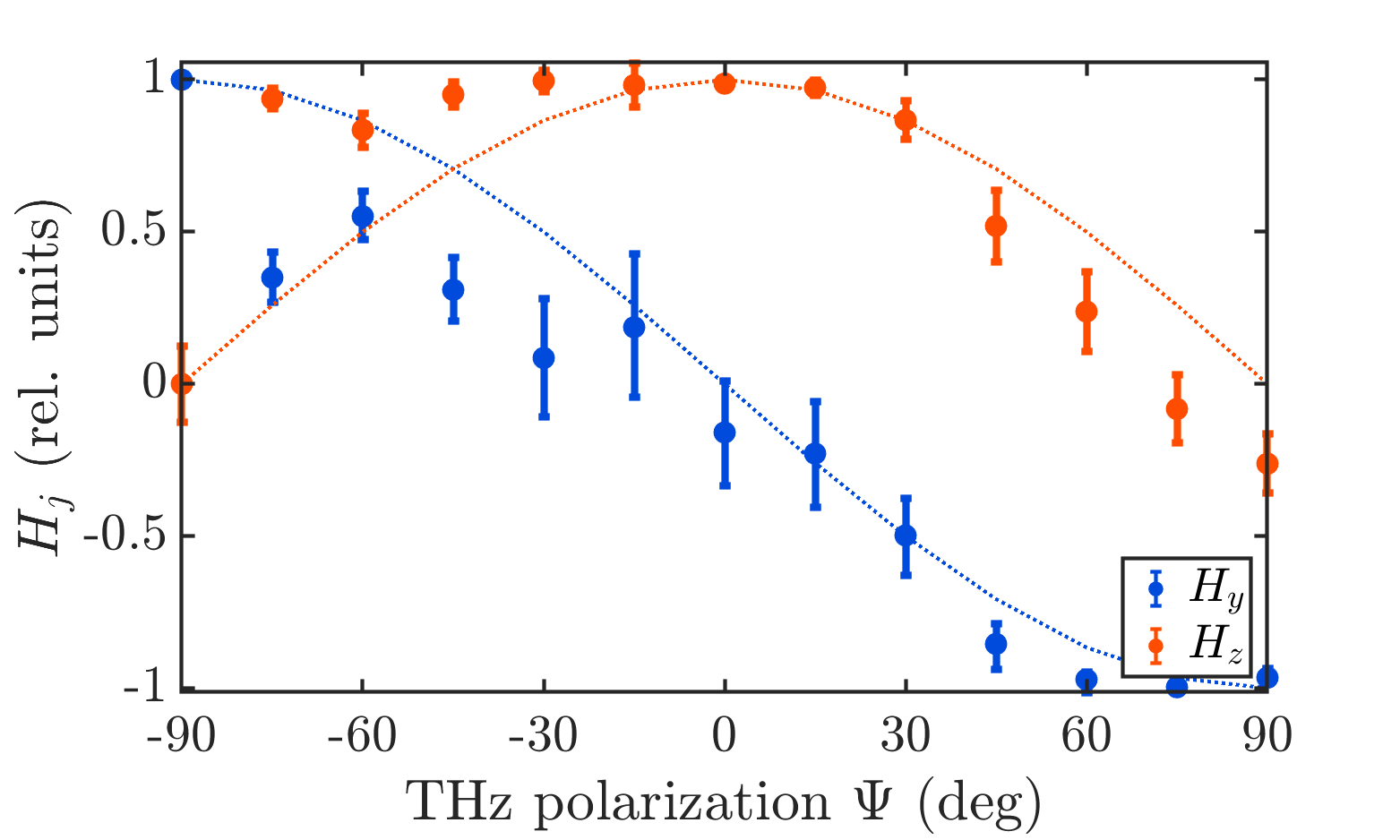}
\caption{{\bf Dependence of spin dynamics in the BIG film covered by the metasurface on polarization of the incident THz pulse.} {\bf a} Spin precession amplitude at $f\approx 0.315$~THz shown by dots was obtained from the Fourier transform of the last part of the experimentally measured precession signal (for $t>45$~ps) with close to the harmonic dynamics. Orange and green colors correspond to $\mathbf{H}_\mathrm{ext} || \mathbf{G}$ and $\mathbf{H}_\mathrm{ext} \perp \mathbf{G}$ orientations (see the insets). Dashed lines depict the theoretical estimations (see the text). {\bf b} Estimation of the ratio of $H_y$ and $H_z$ components launching spin dynamics in BIG film obtained by the relative phase change due to polarization variation for $\mathbf{H}_\mathrm{ext} \perp \mathbf{G}$ orientation (Supplementary, Fig. S7) }\label{Fig: Amplitude from psi}
\end{figure}

Aiming to understand the possibilities for vectorial control of the THz magnetic fields, we studied the THz-induced spin dynamics with the magnetic field of the incident THz pulse continuously rotated in the $xy$ plane from $\Psi=-90^\circ$ to $\Psi=+90^\circ$, where $\Psi$ is the angle between the THz and the external magnetic field. The experiments are performed in two geometries with $\mathbf{H}_\mathrm{ext} \perp \mathbf{G}$ and $\mathbf{H}_\mathrm{ext} || \mathbf{G}$, respectively. The amplitude of the spin oscillations deduced from the dynamic measurements at various  $\Psi$ was plotted in (Fig.~\ref{Fig: Amplitude from psi}) as a function of $\Psi$. 
When  $\mathbf{H}_\mathrm{ext} \perp \mathbf{G}$, only the magnetic field component of the TE-mode can result in a torque and eventually excite the spin resonance. The torque is proportional to  $H_\mathrm{TE,max}^{(BIG)}\sin{\Psi}$, where $H_{TE,max}^{(BIG)}$ is the maximum possible amplitude of the $TE$-component of the THz magnetic field generated in the BIG film in this particular case i.e. when $\Psi=\pm 90^\circ$ (green symbols and curve in Fig.~\ref{Fig: Amplitude from psi}a). 
When $\mathbf{H}_\mathrm{ext} || \mathbf{G}$, both TE- and TM-mode components induce torques on the spins and the total torque is thus proportional to $\sqrt{(H_\mathrm{TM,max}^{(BIG)})^2\sin^2{\Psi}+(H_\mathrm{TE,max}^{(BIG)})^2\cos^2{\Psi}}$, where $H_{TM,max}^{(BIG)}$ is the maximum possible amplitude of the THz-magnetic-field of the $TM$-component in the BIG film in this particular case i.e. when $\Psi=\pm 90^\circ$ (orange symbols and curve in Fig.~\ref{Fig: Amplitude from psi}a).

Apart from the amplitude of the spin precession its phase also varies with the angle $\Psi$: for $\mathbf{H}_\mathrm{ext} \perp \mathbf{G}$     
the phase sharply changes by $180$~deg. when crossing $\Psi=0^\circ$ while for $\mathbf{H}_\mathrm{ext} || \mathbf{G}$ the phase monotonically increases from $-\pi$ to $\pi$ which is due to monotonic relative change between TE- and TM-mode components (Supplementary, Fig. S7).

From the experimental data for the variation of phase one can deduce amplitudes of $\mathbf{H}_y^{BIG}$ and ($\mathbf{H}_z^{BIG}$) components of the THz near-field in the BIG film versus the angle $\Psi$ in the configuration $\mathbf{H}_\mathrm{ext} || \mathbf{G}$ (Fig.~\ref{Fig: Amplitude from psi}b). It shows that varying the relative orientation of the incident THz magnetic field with respect to the metasurface grating, one can tune the ratio of the tangential and normal components of the near-field $\mathbf{H}_\mathrm{THz}$. Indeed, for $\Psi$ close to $\pm\ 90$deg.  the THz field is mainly directed along the metasurface tranches, while for for $\Psi\sim 0$ the THz field has a pronounced out-of-plane ($\mathbf{H}_z^{BIG}$) component. At the intermediate values of $\Psi$ both field components are present. Thus, the metasurface facilitates three dimensional vectorial control of the THz magnetic field allowing, in particular, a gradual change of the direction of the THz near-field from fully in-plane to fully out-of-plane.

%It should be noted that the third component of the THz field ($\mathbf{H}_x$) should also be present but in this configuration it doesn't produce any torque on spins since the latter are aligned along x-axis.     

%The experimentally measured magnitude of the spin precession (see dots in Fig.~\ref{Fig: Amplitude from psi}) is in a good agreement with these theoretical considerations. The values of THz-induced magnetic fields at the resonant wavelength were calculated as $H_{\mathrm{THz}~x}^\mathrm{BIG}=2.9H_0$, $H_{\mathrm{THz}~y}^\mathrm{BIG}=3.0H_0$, $H_{\mathrm{THz}~z}^\mathrm{BIG}=1.8H_0$, where $H_0$ is the magnitude of the incident THz magnetic field. 

\section*{Conclusion}

Here we demonstrate an all-dielectric magnetic metasurface for advanced THz control of spins by ultrashort electromagnetic pulses. It is shown that the magnetic metasurface might not only resonantly amplify THz electromagnetic fields,  but also 'twist' their directions allowing, in particular, vectorial control of the THz magnetic field in all three dimensions. In contrast to plasmonic structures, which may, in principle, be designed to have similar functionalities all-dielectric metasurfaces employ only highly transparent i.e. least dissipative materials. Consequently, such structures are expected to be highly stable against THz pulses of the record high strengths~\cite{ashitkov2024nonequilibrium}, amplify them and thus facilitate application of the record high THz magnetic and electric fields in all three dimensions. These metasurfaces thus offer a new tool for fundamental research in the fields of ultrafast magnetism, photo-induced phase transitions and ultrafast phenomena, THz spintronics and magnonics. 

Generally, metaphotonics approaches are very promising for ultrafast magnetism at sub-THz and THz frequencies since it can provide not only peculiar spin excitation conditions but also an efficient spin control by passing, for example, to a strong spin-photon coupling regime. In this respect, further development of all-dielectric metaphotonic magnetic structures possessing different kinds of THz resonances including guided modes, bound states in the continuum, Mie and whispering gallery modes is highly anticipated.  

\backmatter

\bmhead{Supplementary information}

See supplementary information file 

\bmhead{Acknowledgments}
The work was supported in parts by the Russian Science Foundation (project N 23-62-10024) and the European Research Council ERC Grant Agreement No. 101054664 (SPARTACUS). The
authors declare that this work has been published as a result
of peer-to-peer scientific collaboration between researchers.
The provided affiliations represent the actual addresses of the
authors in agreement with their digital identifier (ORCID)
and cannot be considered as a formal collaboration between
the aforementioned institutions.

%\section*{Declarations}

%Some journals require declarations to be submitted in a standardised format. Please check the Instructions for Authors of the journal to which you are submitting to see if you need to complete this section. If yes, your manuscript must contain the following sections under the heading `Declarations':

%\begin{itemize}
%\item Conflict of interest/Competing interests (check journal-specific guidelines for which heading to use)
%\item Ethics approval and consent to participate
%\item Consent for publication
%\item Data availability 
%\end{itemize}

%\noindent
%If any of the sections are not relevant to your manuscript, please include the heading and write `Not applicable' for that section. 

\section*{Methods}

\subsection*{All-optical pump-probe measurements}\label{All-optical pump-probe}

\subsection*{Metasurface fabrication}\label{App Fabrication}
Figure S1.1 (supplementary) shows the detailed geometry of the metasurface. First, a BIG film of 3.6~$\mu$m thickness was epitaxially grown on a 510~$\mu$m thick GSGG substrate. The metasurface was fabricated on the GSGG substrate by photolithography. Negative photoresist was used to create a protective mask with characteristic dimensions of the structure and to form the topology by chemical etching in orthophosphoric acid. The etching depth was estimated using a profilometer.

\subsection*{THz-pump -- optical-probe technique}\label{App Pumpprobe}
In the pump-probe experimental technique, the laser pulses at the central wavelength of 800 nm and duration of 100 fs were divided into two parts. The most intense one with an energy of 4 mJ  was employed to generate single-cycle  THz pump pulses by the effect of optical rectification of tilted front pulses in a LiNbO$_3$ crystal~\cite{hebling1996derivation}. The THz pulses had an electric field amplitude of 1 MV / cm and a broadband spectrum centered at 0.7 THz (Fig.~\ref{Fig: General}b, gray area and Supplementary) and were focused on the sample using gold-coated parabolic mirrors into a spot with a diameter of 250~$\mu$m. This diameter was measured by moving the probe pulse, utilizing a lens mounted on motorized stages, over a gallium phosphide film. The lower intensity part of the laser pulse was focused on the sample to a beam diameter of 50~$\mu$m and spatially overlapped with the pump to probe spin dynamics by measuring the magneto-optical Faraday effect which is proportional to the oscillating out-of-plane magnetization component. Varying time retardation between the THz pump and optical probe pulse, time-resolved measurements were obtained by measuring probe polarization changes using a balanced photo-detector.  





\bibliography{sn-bibliography}% common bib file

\begin{thebibliography}{10}
\expandafter\ifx\csname url\endcsname\relax
  \def\url#1{\burl{#1}}\fi
\expandafter\ifx\csname urlprefix\endcsname\relax\def\urlprefix{URL }\fi
\providecommand{\bibinfo}[2]{#2}
\providecommand{\eprint}[2][]{\url{#2}}
\providecommand{\doi}[1]{\url{https://doi.org/#1}}
\bibcommenthead

\bibitem{pan2009heat}
\bibinfo{author}{Pan, L.} \& \bibinfo{author}{Bogy, D.~B.}
\newblock \bibinfo{title}{Heat-assisted magnetic recording}.
\newblock \emph{\bibinfo{journal}{Nature Photonics}} \textbf{\bibinfo{volume}{3}}, \bibinfo{pages}{189--190} (\bibinfo{year}{2009}).

\bibitem{zhang2023generation}
\bibinfo{author}{Zhang, Z.} \emph{et~al.}
\newblock \bibinfo{title}{Generation of third-harmonic spin oscillation from strong spin precession induced by terahertz magnetic near fields}.
\newblock \emph{\bibinfo{journal}{Nature Communications}} \textbf{\bibinfo{volume}{14}}, \bibinfo{pages}{1795} (\bibinfo{year}{2023}).

\bibitem{fan2025spatiotemporal}
\bibinfo{author}{Fan, Y.}, \bibinfo{author}{Cao, G.}, \bibinfo{author}{Jiang, S.}, \bibinfo{author}{{\AA}kerman, J.} \& \bibinfo{author}{Weissenrieder, J.}
\newblock \bibinfo{title}{Spatiotemporal observation of surface plasmon polariton mediated ultrafast demagnetization}.
\newblock \emph{\bibinfo{journal}{Nature Communications}} \textbf{\bibinfo{volume}{16}}, \bibinfo{pages}{873} (\bibinfo{year}{2025}).

\bibitem{lee2025enhanced}
\bibinfo{author}{Lee, H.-T.} \emph{et~al.}
\newblock \bibinfo{title}{Enhanced terahertz magneto-plasmonic effect enabled by epsilon-near-zero iron slot antennas}.
\newblock \emph{\bibinfo{journal}{Nanophotonics}}  (\bibinfo{year}{2025}).

\bibitem{schlauderer2019temporal}
\bibinfo{author}{Schlauderer, S.} \emph{et~al.}
\newblock \bibinfo{title}{Temporal and spectral fingerprints of ultrafast all-coherent spin switching}.
\newblock \emph{\bibinfo{journal}{Nature}} \textbf{\bibinfo{volume}{569}}, \bibinfo{pages}{383--387} (\bibinfo{year}{2019}).

\bibitem{haigh2015magneto}
\bibinfo{author}{Haigh, J.} \emph{et~al.}
\newblock \bibinfo{title}{Magneto-optical coupling in whispering-gallery-mode resonators}.
\newblock \emph{\bibinfo{journal}{Physical Review A}} \textbf{\bibinfo{volume}{92}}, \bibinfo{pages}{063845} (\bibinfo{year}{2015}).

\bibitem{almpanis2020spherical}
\bibinfo{author}{Almpanis, E.} \emph{et~al.}
\newblock \bibinfo{title}{Spherical optomagnonic microresonators: triple-resonant photon transitions between zeeman-split mie modes}.
\newblock \emph{\bibinfo{journal}{Physical Review B}} \textbf{\bibinfo{volume}{101}}, \bibinfo{pages}{054412} (\bibinfo{year}{2020}).

\bibitem{rameshti2022cavity}
\bibinfo{author}{Rameshti, B.~Z.} \emph{et~al.}
\newblock \bibinfo{title}{Cavity magnonics}.
\newblock \emph{\bibinfo{journal}{Physics Reports}} \textbf{\bibinfo{volume}{979}}, \bibinfo{pages}{1--61} (\bibinfo{year}{2022}).

\bibitem{pantazopoulos2019high}
\bibinfo{author}{Pantazopoulos, P.~A.}, \bibinfo{author}{Tsakmakidis, K.~L.}, \bibinfo{author}{Almpanis, E.}, \bibinfo{author}{Zouros, G.~P.} \& \bibinfo{author}{Stefanou, N.}
\newblock \bibinfo{title}{High-efficiency triple-resonant inelastic light scattering in planar optomagnonic cavities}.
\newblock \emph{\bibinfo{journal}{New Journal of Physics}} \textbf{\bibinfo{volume}{21}}, \bibinfo{pages}{095001} (\bibinfo{year}{2019}).

\bibitem{klos2014photonic}
\bibinfo{author}{K{\l}os, J.~W.}, \bibinfo{author}{Krawczyk, M.}, \bibinfo{author}{Dadoenkova, Y.~S.}, \bibinfo{author}{Dadoenkova, N.} \& \bibinfo{author}{Lyubchanskii, I.}
\newblock \bibinfo{title}{Photonic-magnonic crystals: Multifunctional periodic structures for magnonic and photonic applications}.
\newblock \emph{\bibinfo{journal}{Journal of Applied Physics}} \textbf{\bibinfo{volume}{115}} (\bibinfo{year}{2014}).

\bibitem{ignatyeva2022all}
\bibinfo{author}{Ignatyeva, D.~O.} \emph{et~al.}
\newblock \bibinfo{title}{All-dielectric magneto-photonic metasurfaces}.
\newblock \emph{\bibinfo{journal}{Journal of Applied Physics}} \textbf{\bibinfo{volume}{132}} (\bibinfo{year}{2022}).

\bibitem{qin2022nanophotonic}
\bibinfo{author}{Qin, J.} \emph{et~al.}
\newblock \bibinfo{title}{Nanophotonic devices based on magneto-optical materials: recent developments and applications}.
\newblock \emph{\bibinfo{journal}{Nanophotonics}} \textbf{\bibinfo{volume}{11}}, \bibinfo{pages}{2639--2659} (\bibinfo{year}{2022}).

\bibitem{bsawmaii2022magnetic}
\bibinfo{author}{Bsawmaii, L.}, \bibinfo{author}{Gamet, E.}, \bibinfo{author}{Neveu, S.}, \bibinfo{author}{Jamon, D.} \& \bibinfo{author}{Royer, F.}
\newblock \bibinfo{title}{Magnetic nanocomposite films with photo-patterned 1d grating on top enable giant magneto-optical intensity effects}.
\newblock \emph{\bibinfo{journal}{Optical Materials Express}} \textbf{\bibinfo{volume}{12}}, \bibinfo{pages}{513--523} (\bibinfo{year}{2022}).

\bibitem{gamet2017enhancement}
\bibinfo{author}{Gamet, E.}, \bibinfo{author}{Varghese, B.}, \bibinfo{author}{Verrier, I.} \& \bibinfo{author}{Royer, F.}
\newblock \bibinfo{title}{Enhancement of magneto-optical effects by a single 1d all dielectric resonant grating}.
\newblock \emph{\bibinfo{journal}{Journal of Physics D: Applied Physics}} \textbf{\bibinfo{volume}{50}}, \bibinfo{pages}{495105} (\bibinfo{year}{2017}).

\bibitem{barsukova2017magneto}
\bibinfo{author}{Barsukova, M.~G.} \emph{et~al.}
\newblock \bibinfo{title}{Magneto-optical response enhanced by mie resonances in nanoantennas}.
\newblock \emph{\bibinfo{journal}{Acs Photonics}} \textbf{\bibinfo{volume}{4}}, \bibinfo{pages}{2390--2395} (\bibinfo{year}{2017}).

\bibitem{ignatyeva2020all}
\bibinfo{author}{Ignatyeva, D.~O.} \emph{et~al.}
\newblock \bibinfo{title}{All-dielectric magnetic metasurface for advanced light control in dual polarizations combined with high-q resonances}.
\newblock \emph{\bibinfo{journal}{Nature communications}} \textbf{\bibinfo{volume}{11}}, \bibinfo{pages}{5487} (\bibinfo{year}{2020}).

\bibitem{xia2022circular}
\bibinfo{author}{Xia, S.} \emph{et~al.}
\newblock \bibinfo{title}{Circular displacement current induced anomalous magneto-optical effects in high index mie resonators}.
\newblock \emph{\bibinfo{journal}{Laser \& Photonics Reviews}} \textbf{\bibinfo{volume}{16}}, \bibinfo{pages}{2200067} (\bibinfo{year}{2022}).

\bibitem{chernov2020all}
\bibinfo{author}{Chernov, A.~I.} \emph{et~al.}
\newblock \bibinfo{title}{All-dielectric nanophotonics enables tunable excitation of the exchange spin waves}.
\newblock \emph{\bibinfo{journal}{Nano letters}} \textbf{\bibinfo{volume}{20}}, \bibinfo{pages}{5259--5266} (\bibinfo{year}{2020}).

\bibitem{ignatyeva2024optical}
\bibinfo{author}{Ignatyeva, D.~O.} \emph{et~al.}
\newblock \bibinfo{title}{Optical excitation of multiple standing spin modes in three-dimensional optomagnonic nanocavities}.
\newblock \emph{\bibinfo{journal}{Physical Review Applied}} \textbf{\bibinfo{volume}{21}}, \bibinfo{pages}{034017} (\bibinfo{year}{2024}).

\bibitem{walowski2016perspective}
\bibinfo{author}{Walowski, J.} \& \bibinfo{author}{M{\"u}nzenberg, M.}
\newblock \bibinfo{title}{Perspective: Ultrafast magnetism and thz spintronics}.
\newblock \emph{\bibinfo{journal}{Journal of Applied Physics}} \textbf{\bibinfo{volume}{120}} (\bibinfo{year}{2016}).

\bibitem{blank2021thz}
\bibinfo{author}{Blank, T.~G.} \emph{et~al.}
\newblock \bibinfo{title}{Thz-scale field-induced spin dynamics in ferrimagnetic iron garnets}.
\newblock \emph{\bibinfo{journal}{Physical Review Letters}} \textbf{\bibinfo{volume}{127}}, \bibinfo{pages}{037203} (\bibinfo{year}{2021}).

\bibitem{parchenko2016non}
\bibinfo{author}{Parchenko, S.} \emph{et~al.}
\newblock \bibinfo{title}{Non-thermal optical excitation of terahertz-spin precession in a magneto-optical insulator}.
\newblock \emph{\bibinfo{journal}{Applied Physics Letters}} \textbf{\bibinfo{volume}{108}} (\bibinfo{year}{2016}).

\bibitem{grishunin2019transient}
\bibinfo{author}{Grishunin, K.} \emph{et~al.}
\newblock \bibinfo{title}{Transient second harmonic generation induced by single cycle thz pulses in ba0. 8sr0. 2tio3/mgo}.
\newblock \emph{\bibinfo{journal}{Scientific reports}} \textbf{\bibinfo{volume}{9}}, \bibinfo{pages}{697} (\bibinfo{year}{2019}).

\bibitem{subkhangulov2016terahertz}
\bibinfo{author}{Subkhangulov, R.} \emph{et~al.}
\newblock \bibinfo{title}{Terahertz modulation of the faraday rotation by laser pulses via the optical kerr effect}.
\newblock \emph{\bibinfo{journal}{Nature Photonics}} \textbf{\bibinfo{volume}{10}}, \bibinfo{pages}{111--114} (\bibinfo{year}{2016}).

\bibitem{ashitkov2024nonequilibrium}
\bibinfo{author}{Ashitkov, S.} \emph{et~al.}
\newblock \bibinfo{title}{Nonequilibrium heating of electrons, melting, and modification of a nickel nanofilm by an ultrashort terahertz pulse}.
\newblock \emph{\bibinfo{journal}{JETP Letters}} \textbf{\bibinfo{volume}{120}}, \bibinfo{pages}{580--588} (\bibinfo{year}{2024}).

\bibitem{hebling1996derivation}
\bibinfo{author}{Hebling, J.}
\newblock \bibinfo{title}{Derivation of the pulse front tilt caused by angular dispersion}.
\newblock \emph{\bibinfo{journal}{Optical and Quantum Electronics}} \textbf{\bibinfo{volume}{28}}, \bibinfo{pages}{1759--1763} (\bibinfo{year}{1996}).

\end{thebibliography}

\end{document}